\documentclass[twocolumn,preprintnumbers,amsmath,amssymb]{revtex4}

\usepackage{graphicx}
\usepackage{dcolumn}
\usepackage{bm}

\begin{document}

\title{Spin Flavor Oscillation of Neutrinos
  in Rotating Gravitational Fields and Their Effects on Pulsar Kicks}

\author{G. Lambiase}
\email{lambiase@sa.infn.it}

\affiliation{Dipartimento di Fisica E.R. Caianiello, Universita'
di Salerno, 84081 - Baronissi (SA), Italy,\\}

\affiliation{INFN, Gruppo collegato di Salerno, Italy.}

\date{\today}

\begin{abstract}

The origin of high velocities of pulsars is studied by considering
the spin-flip conversion of neutrinos propagating in a
gravitational field of a protoneutron star. For a rotating
gravitational source (such as pulsars) with angular velocity ${\bm
\omega}$, one finds that the spin connections (entering in the
Dirac equation written in curved space time) induce an additional
contribution to neutrino energy which is proportional to ${\bm
\omega}\cdot {\bf p}$, with ${\bf p}$ the neutrino momentum. Such
a coupling (spin-gravity coupling) can be responsible of pulsar
kicks being the asymmetry of the neutrino emission generated by
the relative orientation of the neutrino momentum ${\bf p}$ with
respect to the angular velocity ${\bm \omega}$. As a consequence,
the mechanism suggests that the motion of pulsars is correlated to
their angular velocity ${\bm \omega}$. In this work we consider
neutrinos propagating orthogonally to the magnetic field. The
fractional asymmetry turns out to be independent on the magnetic
field of the nascent protostar, and is only related to the angular
velocity ($\Delta p/p \sim \omega$). As in the usual approaches,
spin flip conversion is generated via the coupling of the neutrino
magnetic momentum with the magnetic field. For our estimations, we
use the large non-standard neutrino magnetic momentum provided by
astrophysical and cosmological constraints, $\mu_\nu\sim
10^{-11}\mu_B$.

The connection with recent observations and statistical analysis
is also discussed.
\end{abstract}

\maketitle

\section{Introduction}

A discussed and unsolved issue of the modern astrophysics is the
origin of the pulsar velocity, i.e. the high proper velocities of
pulsars as compared with the surrounding stars. Their
three-dimensional galactic speed runs, in fact, from $450\pm
90$Km/sec up to a maximum of about 1000Km/sec \cite{Lyne}. This
peculiarity immediately has suggested that nascent pulsars receive
an impulse (kick) during their formation. The gravitational
binding energy ($3\times 10^{53}$erg) is, after the supernova
collapse of a massive star, carried out by outgoing neutrinos
(about $99\%$). An anisotropy of $\sim 1\%$ of the momenta
distribution of the outgoing neutrinos would then suffice to
account for the neutron star recoil of 300Km/sec.

An interesting mechanism to generate the pulsar velocity has been
recently proposed by Kusenko and Segr\'e (KS) \cite{kusenkoPRL}.
It involves the physics of neutrino oscillation in presence of an
intense magnetic field. Let us recall the basic idea. The
neutrinosphere is defined as the surface from which neutrinos may
escape from the protostars. In particular, electron neutrinos
$\nu_e$ are emitted from a surface which is located at a distance
from the center greater than the surfaces corresponding to
muon/tau ($\nu_{\mu, \tau}$) neutrinospehres. Under suitable
conditions, a resonant oscillation $\nu_e\to \nu_{\mu,\tau}$ can
occur between the $\nu_e$ and $\nu_{\mu,\tau}$ neutrinospheres.
Neutrinos $\nu_{\mu,\tau}$ generate via oscillations can escape
from the protostar being outside of their neutrinosphere, with the
ensuing that the "surface of the resonance" acts as an "effective
$\nu_{\mu, \tau}$-neutrinosphere". The presence of a magnetic
field may distort the effective surface of resonance and the
energy flux turns out to be generated anisotropically. In the KS
mechanism \cite{kusenkoPRL}, the responsible for the neutrino
emission anisotropy is the polarization of the medium due to the
magnetic field ${\bf B}$. The usual MSW resonance conditions turn
out to be, in fact, modified by the term (first derived by
D'Olivo, Nieves and Pal) \cite{pal}
 \[
\frac{eG_F}{\sqrt{2}}\left(\frac{3n_e}{\pi^4}\right)^{1/3}{\bf
B}\cdot {\hat {\bf p}}
 \]
where ${\hat {\bf p}}={\bf p}/p$, ${\bf p}$ is the neutrino
momentum, $e$ is the electric charge, $G_F$ is the Fermi constant,
and $n_e$ is the electron density. The KS mechanism has been also
studied for active-sterile neutrino oscillations (sterile
neutrinos may have a small mixing angle with the ordinary
neutrinos) \cite{kusenkoPLB,fuller}. Papers dealing with the
origin of pulsar kicks can be found in
\cite{kusenkoPRD99,qian,lanza,grasso,horvat,zanella,casini,nardi,burrows,cuesta,loveridge,goyal,lai,all,janka,lambiasePRL,lambiasePRD,voloshin,chugai,dorofeev}.

In this paper is discussed the possibility that spin flavor
conversion of neutrinos propagating in a gravitational field
generated by a rotating source may generate the observed pulsar
kicks \cite{lambiasePRL} (the role of the angular velocity on
pulsar kicks has been also studied, although in a different
context, by Mosquera Cuesta \cite{cuesta}). The gravitational
field affects, as we will see, the resonance conversion of
left-handed neutrinos into right handed neutrinos, the latter
being sterile can escape from the neutrinosphere, hence from the
protoneutron star. In particular, the modification to the
resonance condition is induced by spin connections which enter in
the Dirac equation in curved spacetimes. They give rise to a
coupling term $\sim {\bm \omega}\cdot {\bf p}$, where ${\bm
\omega}$ is the angular velocity of the gravitational source. The
relative orientation of neutrino momenta with respect to the
angular velocity determines an asymmetry in the neutrino emission,
hence may generate pulsar kicks.

The paper is organized as follows. In Sect. II we review the Dirac
equation in curved space-times. Here we see that, owing to the
breakdown of the spherical symmetry generated by the angular
velocity of the gravitational source, spin connections are non
vanishing and are proportional to the chiral operator $\gamma^5$.
Sect. III and Sect. IV are devoted, respectively, to briefly
recall the main features of matter induced effective potential and
the role and intensity of magnetic fields in astrophysical
systems. In Sect. V the fractional asymmetry is computed. Here we
also discuss the resonance and adiabatic conditions, as well as
the spin flip probability that left-handed neutrinos transform in
right-handed neutrinos. Conclusions are drawn in Sect. VI.

\section{Dirac Equation in Curved Space Time}

The phase of neutrinos propagating in a curved background is
generalized as \cite{CAR,altri}
\begin{equation}\label{1}
\vert\psi_{f}(\lambda)\rangle = \sum_{j} U_{f j}\,
e^{i\int_{\lambda_0}^{\lambda}P\cdot
p_{null}d\lambda^{\prime}}\vert\nu_j\rangle\,{,}
\end{equation}
where $f$ is the flavor index and $j$ the mass index. $U_{f j}$
are the matrix elements transforming flavor and mass bases
 \begin{equation}\label{U}
U =\left(\begin{array}{cc}
                \cos\theta & \sin\theta \\
                 -\sin\theta & \cos\theta \end{array}\right)\,,
 \end{equation}
(in what follows we shall consider the neutrino mixing for two
flavors). Besides,
 \[
 P\cdot p_{null}=P_\mu \, p^\mu_{null}\,,
 \]
where $P_\mu$ is the four--momentum operator generating
space--time translation of the eigenstates and
 \[
 p^{\mu}_{null}=\frac{dx^{\mu}}{d\lambda}\,,
 \]
is the tangent vector to the neutrino worldline $x^{\mu}$,
parameterized by $\lambda$. From here we follow the Cardall and
Fuller paper \cite{CAR}. The covariant Dirac equation in curved
space--time is (in natural units) \cite{WEI}
 \[
[i\gamma^{\mu}(x)D_{\mu}-m]\psi=0\,,
 \]
where the matrices $\gamma^{\mu}(x)$ are related to the usual
Dirac matrices $\gamma^{\hat{a}}$ by means of the vierbein fields
$e^{\mu}_{\hat{a}}(x)$, i.e.
 \[
\gamma^{\mu}(x)=e^{\mu}_{\hat{a}}(x)\gamma^{\hat{a}}\,.
 \]
The Greek (Latin with hat) indices refer to curved (flat)
space--time. $D_{\mu}$ is defined as
 \[
 D_{\mu}=\partial_{\mu}+\Gamma_{\mu}(x)\,,
 \]
where $\Gamma_{\mu}(x)$ are the spin connections
 \[
\Gamma_{\mu}(x)=\frac{1}{8} [\gamma^{\hat{a}},
\gamma^{\hat{b}}]e^{\nu}_{\hat{a}}e_{\nu \hat{b};\mu}\,,
 \]
(semicolon represents the covariant derivative). Using the
relation
\[
\gamma^{\hat a}[\gamma^{\hat b}, \gamma^{\hat c}]= 2\eta^{{\hat
a}{\hat b}}\gamma^{\hat c}-2\eta^{{\hat a}{\hat c}}\gamma^{\hat b}
-2i\epsilon^{{\hat d}{\hat a}{\hat b}{\hat c}}\gamma_{\hat
d}\gamma^5
\]
where $\eta_{{\hat a}{\hat b}}$ is the metric of flat spacetime,
and after some manipulations, the spin connections can be cast in
the form \cite{CAR}
 \[
\gamma^\mu\Gamma_{\mu}=\gamma^{\hat{a}}e^{\mu}_{\hat{a}}\Gamma_{\mu}=
\gamma^{\hat{a}}e^{\mu}_{\hat{a}}
\left\{iA_{G\mu}\left[-(-g)^{-1/2}\, \gamma^5\right]\right\}\,,
 \]
where
 \[
A_G^{\mu}=\frac{1}{4}\sqrt{-g}e^{\mu}_{\hat{a}}\varepsilon^{\hat{a}\hat{b}\hat{c}\hat{d}}
(e_{\hat{b}\nu;\sigma}-e_{\hat{b}\sigma;\nu})e^{\nu}_{\hat{c}}e^{\sigma}_{\hat{d}}\,,
 \]
and
 \[
g=\det(g_{\mu\nu})\,.
 \]
The above procedure allows to separate out the chirality operator
$\gamma^5$. This shows that $\gamma^\mu \Gamma_\mu$ acts
differently on left- and right-handed neutrino states. In fact, by
writing
 \[
 \gamma^5 = P_R-P_L\,,
 \]
where
 \[
P_{L,R}=\frac{1\mp \gamma^5}{2}
 \]
are the projection operators, one sees that neutrinos with left-
and right-handed helicity acquire a different gravitational
contribution. In the case of neutrino oscillations, one can add,
without physical consequences, a term proportional to the identity
matrix ($\sim A_{G\mu} I$), so that $\gamma^5$ can be replaced by
the left-handed projection operator ${\cal P}_L=(1-\gamma^5)/2$.
As a consequence, the spin-gravity coupling is pushed in the
left-handed sector of the effective Hamiltonian (see Eqs.
(\ref{12})-(\ref{12b})), and no contributions appear in the right
handed sector.

The equation of evolution of neutrinos has the form ($f=e$,
$f'=\mu, \tau$)
 \[
 i\frac{d}{d\lambda}\left(\begin{array}{c}
                           \nu_{fL} \\
                             \nu_{f' L} \\
                              \nu_{fR} \\
                           \nu_{f' R}\end{array}\right)={\cal H}\left(\begin{array}{c}
                                                            \nu_{fL} \\
                             \nu_{f' L} \\
                              \nu_{fR} \\
                           \nu_{f' R}\end{array}\right)\,,
 \]
where the diagonal terms of the Hamiltonian ${\cal H}$
 \[
 \mbox{diag} {\cal H}=U^{-1}MU+\Omega_G(x)\, {\cal P}_L\,,
 \]
are written in terms of the mixing matrix $U$ and mass matrix $M$
\[
  M= \left(\begin{array}{cc}
  m_1^2 & 0 \\
  0 & m_2^2 \end{array}\right)\,,
\]
and $\Omega_G$ is defined as
\[
 \Omega_G(x)\equiv \frac{p^\mu A_{G\,\mu}}{E}\,.
\]
$m_1$ and $m_2$ are the mass eigenstates, $\theta$ is the vacuum
mixing angle, and $p^\mu=(E, {\bf p})$, being $E$ the energy
measured in the local frame.

The inclusion of matter induced effective potential and magnetic
terms, which appear in the off-diagonal terms of the Hamiltonian
${\cal H}$, will be discussed later.

\subsection{The Geometry of a Rotating Mass Source}

For geometries with a spherical symmetry, such as the
Schwarzschild or Reissner-Nordstrom space-times, $A_{G}^{\mu}$
vanish. Nevertheless, for rotating gravitational sources
(Lense-Thirring geometry), whose line element is (in weak field
approximation)
 \begin{equation}\label{9}
 ds^2=(1-\phi)(dt)^2-(1+\phi)(d{\bf x})^2
-2{\bf h}\cdot d{\bf x}\,dt \,,
 \end{equation}
with
 \[
{\bf x}=(x, y, z)\,, \quad
 r=\sqrt{x^2+y^2+z^2}\,,
 \]
 \[
\phi(r)=\frac{2GM}{r}\,,\quad {\bf
h}=\left(\frac{4GMR^2}{5r^3}\right)\,{\bm \omega}\wedge {\bf x}\,,
 \]
${\bm \omega}$ is the angular velocity of the gravitational mass
$M$, and $R$ its radius, $A_G^\mu$ acquire a non vanishing
component related to the off-diagonal terms of the metric tensor,
 \[
 A_G^\mu(x)=\left(0, -\frac{4}{5}\frac{GM R^2}{r^3}\, {\bm
\omega}^\prime\right)\,,
 \]
where
 \[
{\bm \omega}^\prime={\bm \omega}-\frac{3({\bm \omega}\cdot {\bf x
}){\bf x}}{r^2}\,.
 \]
One can show the angular velocity induces a {\it drift} velocity
of neutrinos \cite{casinidrift}.

The vierbein fields used for computing the spin connection
$A_{G}^\mu$ are
 \[
 e_{0\, {\hat 0}}=(1+\phi)\,, \quad e_{j\,{\hat i}}=-(1-\phi)\delta_{ji}\,,
 \]
 \[
 e_{0\, {\hat i}}=-h^i\,, \quad e_{i\,{\hat 0}}=0 \,.
 \]
The non vanishing $A_G^\mu$ is an indication of a preferred
direction related to the angular velocity of the source.
$\Omega_G(x)$ can be rewritten as
\begin{eqnarray}\label{omegaG}
  \Omega_G(r)&=&\frac{p_\mu A^{\mu}_G(r)}{E_l}
  =\frac{4GMR^2}{5 r^3E_l}\,
  {\bf p}\cdot {\bm \omega}'\, \\
 & \sim & 10^{-13}
  \frac{M}{M_\odot}\left(\frac{R}{10\mbox{km}}\right)^2
  \left(\frac{20\mbox{km}}r \right)^3\frac{\omega' \cos \beta}{10^4\mbox{Hz}}
  \, \mbox{eV}\,. \nonumber
\end{eqnarray}
As the angular velocity goes to zero, the spherical symmetry is
recovered and the spin connections vanish, as immediately follows
from the above expressions.

It is worth note that the term $\gamma^\mu\Gamma_\mu$ can be
rewritten in the form ($\sqrt{-g}\sim 1+2\phi$)
 \[
 i\gamma^\mu\Gamma_\mu \simeq i\gamma^\mu (-iA_{G\,
 \mu}\gamma^5)=f(r)\gamma^0\omega^{\prime \,i}\Sigma^i
 \]
where
 \[
 f(r)=\frac{4GMR^2}{5r^3}\,,
 \]
and
 \[
 \Sigma^i=\gamma^5 \gamma^0\gamma^i=\left(\begin{array}{cc}
                \sigma^i & 0 \\
                 0 & \sigma^i \end{array}\right)\,.
 \]
Here $\sigma^i$, $i=1, 2, 3$ are the Pauli matrices.

The term $\sim f(r) {\bm \omega}'\cdot {\bm \Sigma}$ is the well
known gravitomagnetic-spin coupling (see for example
\cite{ciufolini}).

\section{Matter Induced Effective Potential}

Neutrinos inside their neutrinospheres are trapped owing to weak
interactions with the background matter, which lead to the
potential energy
 \[
V_{\nu_{f}}\simeq 3.8\frac{\rho}{10^{14}\mbox{gr
cm}^{-3}}y_{f}(r,t)\mbox{eV}\,,
 \]
where $\rho=m_n n_e$ is the matter density, $m_n$ is the nucleon
mass, $f=e, \mu, \tau$, and
 \[
y_e=Y_e-1/3\,, \quad y_{\mu, \tau}=Y_e-1\,.
 \]
In these expressions, $Y_e$ is the electron fraction. In the
present analysis we shall envisage those neutrinos for which
matter induced effective potential for (electron) left-handed
neutrinos such that $V_{\nu_e}\ll 1$. This follows in the regions
where the electron fraction $Y_e$ assumes the value $\approx 1/3$
($y_f\ll1$) \cite{pons,voloshin,valle}. These regions are located
at $r\sim 15$ km (see the paper by Nunokawa, Peltoniemi, Rossi and
Valle \cite{valle}).

\section{Magnetic Field Inside a Protostar}

Electroweak interactions of neutrinos with matter background play
a central role on the neutrino emission during the core collapse
of supernovae, and in general, on star cooling mechanism of
magnetized medium. The neutrino energy spectrum in presence of
strong magnetic fields is modified, depending on their flavors.

Protoneutron stars possess strong magnetic fields whose strength
is $\gtrsim 10^{12} - 10^{14}$G. Such strong fields can be also
found near the surface of supernovae \cite{BSN}, neutron stars
\cite{Bneutronstar}, and magnetostars \cite{Bmagnetostar}. A
fundamental feature of large magnetic fields in such astrophysical
systems is related to their effects on neutrinos, which through
charged and neutral current interactions, modify the dispersion
relations of neutrinos.

In computing the effects of strong magnetic fields on neutrinos
propagation, the following condition holds
 \[
 m_e^2 \ll eB \ll M_W^2,
 \]
where $m_e$ is the electron mass and $M_W$ is the $W$-boson mass.
We define the fields
 \[
 B_c=\frac{m_e^2}{e}\sim 4.4\times 10^{13}\mbox{G}\simeq
 8.62 \times 10^{11}\mbox{eV}^2\,.
 \]
It "separates" the regimes of weak field $B\ll B_c$,  and of
strong field $B\gtrsim B_c$.

In a series of recent papers \cite{erdas,elizalde}, a detailed
analysis of the neutrinos physics in strong magnetic fields has
been carried out. In what follows we shall consider those
neutrinos which propagate orthogonally to the magnetic field of
the protostar, so that ${\bf B}\cdot {\hat {\bf p}}=B\cos\alpha=0$
($\alpha=\pi/2$). The only contribution to the neutrino energy
comes from the magnetic momentum of neutrinos.

\subsection{Neutrino Magnetic Momentum}

Since neutrinos are uncharged particles, they do not interact
directly with photons (magnetic fields). The typical coupling of
the electromagnetic field with the fermionic current is absent for
neutrinos. This is essentially due to the fact that the Standard
Model is built up assuming that neutrinos are massless, hence only
left-handed neutrinos appear in the theory. The experimental
evidence of neutrino oscillations is an index that neutrinos are
massive particles. In such a circumstance, we are in the context
of a theory which goes beyond the Standard Model since
right-handed projection of neutrinos has to be included in the
fermion sector. As a consequence, an anomalous magnetic momentum
emerges through quantum corrections (one-loop diagram whose
internal lines are charged lepton and $W$-$Z$ boson propagators).
Thus, even though neutrinos have no charged, they posses a
magnetic momentum, which induces the interaction with photons
\cite{muSM}
 \[
{\hat \mu}_\nu=\mu_{ff'}=\frac{3eG_Fm_\nu}{8\sqrt{2\pi^2}} \sim
10^{-19}\mu_B\frac{m_\nu}{\mbox{eV}} \,,
 \]
where $m_\nu$ is the neutrino mass, and
 \[
\mu_B=\frac{e}{2m_e}
 \]
is the Bohr magneton.

Neutrinos interacting with the magnetic field of the protoneutron
star acquire, hence, an energy through the interaction \cite{okun}
 \[
{\cal L}_{int}={\bar \psi}{\hat \mu}_\nu\sigma^{{\hat a}{\hat
b}}F_{{\hat a}{\hat b}}\psi\,,
 \]
where  $F_{{\hat a}{\hat b}}$ is the electro-magnetic field
tensor, and
 \[
\sigma^{{\hat a}{\hat b}}=\frac{1}{4}[\gamma^{\hat a},
\gamma^{\hat b}]\,.
 \]
In this paper, we shall consider the neutrino magnetic momentum
 \[
 \mu_{ff'}\sim (10^{-12}-10^{-11})\mu_B\,,
 \]
as provided by astrophysical and cosmological constraints
\cite{pakvasa}. Such a value on the neutrino magnetic momentum is
great as compared with ones of the Standard Model prediction
$\mu_{ff'}\sim 10^{-19}\mu_B$, provided that the neutrino mass is
$m_\nu\sim$ few eV \cite{muSM}.

The expression for the neutrino magnetic energy is
 \[
\mu_{ff'}B\sim 2.5\times
10^{-6}\frac{\mu_{ff'}}{10^{-11}\mu_B}\frac{B}{B_c}\, \mbox{eV}\,.
 \]

\section{Equation of evolution of neutrinos and the Asymmetric Neutrino Emission}

Taking into account for the gravitational and magnetic
contributions, the equation of evolution describing the conversion
between two neutrino flavors $f=e$ and $f'=\mu, \tau$ reads
\cite{PIN}
 \begin{equation}\label{11}
i\frac{d}{d\lambda}\left(\begin{array}{c}
                           \nu_{fL} \\
                             \nu_{f' L} \\
                              \nu_{fR} \\
                           \nu_{f' R}\end{array}\right)={\cal H}\left(\begin{array}{c}
                                                            \nu_{fL} \\
                             \nu_{f' L} \\
                              \nu_{fR} \\
                           \nu_{f' R}\end{array}\right)\,,
 \end{equation}
where, in the chiral base, the matrix ${\cal H}$ is the effective
Hamiltonian defined as
\begin{equation}\label{12}
{\cal H}=\left[\begin{array}{cc}
 {\cal H}_L & {\cal H}_{ff'}^* \vspace{0.05in} \\
 {\cal H}_{ff'} & {\cal H}_R \\
\end{array}\right]\,,
\end{equation}
\begin{equation}\label{12a}
{\cal H}_L=\left[\begin{array}{cc}
 V_{\nu_e}+\Omega_G-\delta c_2
         &  \delta s_2 \vspace{0.05in} \\
 \delta s_2  & V_{\nu_{f'}}+\Omega_G +\delta c_2 \\
\end{array}\right],
\end{equation}
\begin{equation}\label{12b}
{\cal H}_R=\left[\begin{array}{cc}
 -\delta c_2 &  \delta s_2 \vspace{0.05in} \\
 \delta s_2  & \delta c_2 \\
\end{array}\right],
 \end{equation}
 \[
{\cal H}_{ff'}=B_\perp \left[\begin{array}{cc}
 \mu_{ff} & \mu_{ff'} \vspace{0.05in} \\
 \mu_{ff'} & \mu_{f'f'} \\
\end{array}\right].
 \]
 \[
\delta=\frac{\Delta m^2}{4E}\,, \quad  \Delta m^2=m_2^2-m_1^2\,,
 \]
 \[
 c_2=\cos 2\theta\,, \quad s_2=\sin 2\theta\,,
 \]
$B_\perp=B\sin\alpha=B$ is the component of the magnetic field
orthogonal to the neutrino momentum, and $\beta$ is the angle
between the neutrino momentum and the angular velocity.

$\Omega_G$ is diagonal in spin space, so that it cannot induce
spin-flips\footnote{Of course $\Omega_G$ does not induce neutrino
oscillations, unless one does not assume a violation of the
equivalence principle, as in Refs \cite{gasp}.}. Its relevance
comes from the fact that it modifies the resonance conditions of
spin-flips
\begin{eqnarray}
 \nu_{fL}\to \nu_{f' R} & V_{\nu_e}+\Omega_G({\bar r})-
 2\delta c_2=0\,,
 \label{res1} \\
   \nu_{f' L}\to \nu_{fR} & \quad
 V_{\nu_{f'}}+\Omega_G({\bar r})+2\delta c_2=0\,.
 \label{res2}
\end{eqnarray}
${\bar r}$ is the radial distance where the resonance occurs. The
resonances (\ref{res1}) and (\ref{res2}) do not occur
simultaneously; in what follows we shall consider the transition
(\ref{res1}).

Besides, we shall use the best fit for solar neutrinos
\cite{kamLAND}
 \begin{equation}\label{bestfit}
\Delta m^2_{Sun}\sim (10^{-5}\div 10^{-4})\mbox{eV}^2\,, \quad
\sin^2 2 \theta_{Sun}\gtrsim 0.8 \,.
\end{equation}

$\Omega_G$ in (\ref{res1}) (and (\ref{res2})) distorts the surface
of resonance due to the relative orientation of the neutrino
momentum with respect to the angular velocity. As a consequence,
the outgoing energy flux ${\bf F}$ results modified. The neutrino
momentum asymmetry is defined as \cite{zanella}
\begin{equation}\label{flux1}
  \frac{\Delta p}{p}=\frac{1}{3}\frac{\int_0^\pi {\bf F}\cdot {\hat {\bm \omega}}da}
  {\int_0^\pi {\bf F}\cdot {\hat {\bf n}}da}\,,
\end{equation}
where the factor 1/3 accounts for the conversion of neutrinos
$\nu_{fL}$ into $\nu_{f'R}$ ($f\neq f'$), $da$ is the element of
area on the distorted surface, ${\hat {\bm \omega}}$ and ${\hat
{\bf n}}$ are the unit vectors parallel to the angular velocity
and orthogonal to $da$, respectively. To compute the fractional
asymmetry one then should specify the protostar model
\cite{shapiro}, including into the hydrodynamical equations the
rotational effects due to gravitational sources \cite{lambiase1}.
This task goes beyond the aim of this work.

Here we give an estimation of the fractional asymmetry following
the calculations as in KS \cite{kusenkoPRL}. The surface of
resonance is located to radial distance $r(\beta)={\bar r}+\varrho
\cos\beta$ ($\cos\beta= {\hat {\bf p}}\cdot {\hat {\bm \omega}}'$,
with ${\hat {\bm \omega}}'={\bm \omega}'/\omega'$). Inserted in
the resonance equation, one has
\begin{equation}\label{deviation}
  \varrho = -\frac{f(r)}{V_{\nu_e}'+\Omega_G'}\simeq
  -\frac{f(r)}{V_{\nu_e}'}\,,
\end{equation}
where we used the resonance condition
\begin{equation}\label{res1a}
  2\delta c_2 = V_{\nu_e}(r)\,,
\end{equation}
and $\Omega_G'\ll V_{\nu_e}'$ as evaluated at ${\bar r}$. The
prime means derivative with respect to $r$. In such a calculations
we assumed that the matter density profile is described by
\cite{raffeltApJ}
 \[
 \rho(r) = \rho_0\left(\frac{15\mbox{km}}{r}\right)^p\,,
 \]
where $\rho_0=2\,\, 10^{14}$gr/cm$^3$ and $p={\cal O}(1)$. The
fractional asymmetry reads
\begin{eqnarray}\label{fracasymmetry}
  \frac{\Delta
  p}{p}&=&\frac{2\varrho}{3T}\frac{dT}{dr}=\frac{2}{3}\frac{f({\bar
  r})}{T}\left(\frac{dV_{\nu_e}}{dT}\right)^{-1} \\
   &=& 82.5 \,\, \frac{\mbox{MeV}^3}{T\, y_e}\,\frac{f({\bar r})}{\mbox{eV}}
   \left(\frac{dn_e}{dT}\right)^{-1}\,. \nonumber
\end{eqnarray}
Since the density number $n_e$ is related to the temperature $T$
via the Fermi distribution
\begin{equation}\label{Fermi}
  n_e= 2\int \frac{d^3p}{(2\pi)^3}\frac{1}{e^{(p-\mu)/T}+1}\,,
\end{equation}
where $\mu$ is the chemical potential, one obtains $dn_e/dT\sim
2\eta T^2/3$, where $\eta \equiv \mu/T\sim 5$ (see the paper by
Qian \cite{qian}). Eq. (\ref{fracasymmetry}) then becomes
\begin{equation}\label{asymmfinal}
  \frac{\Delta p}{p}\sim 0.01 \left[4.5 \,\, 10^5\left(
  \frac{3\mbox{MeV}}{T}\right)^3 \frac{1}{\eta \, y_e}\,
  \frac{f({\bar r})}{\mbox{eV}}\right]\,.
\end{equation}
From the resonance condition (\ref{res1}) one obtains
 \[
\Delta m^2 \cos 2 \theta \simeq 7.6 \,\, 10^{4} y_e\, \mbox{eV}^2,
 \]
as the density $\rho\sim 10^{11}$gr/cm$^3$ at ${\bar r}\sim 15$km.
The value falls into the best fit (\ref{bestfit}) provided
$y_e\approx 10^{-9}$. One can then see that for typical values of
pulsars
 \[
R\gtrsim 15\mbox{km}\,, \quad M\sim M_\odot\,, \quad T\sim
3\mbox{MeV}\,,
 \]
Eq. (\ref{asymmfinal}) implies that the fractional asymmetry $\sim
1\%$ is recovered as
 \[
\omega \sim \mbox{few}\,\,10^2\mbox{Hz}\,,
 \]
which is, indeed, the typical angular velocity of pulsars.

\subsection{Adiabatic Conditions}

Besides the level crossing (\ref{res1}), it must be also
adiabatic, i.e. the corresponding (adiabatic) parameter $\gamma$,
which quantifies the magnitude of the off-diagonal elements with
respect to the diagonal ones of (\ref{12}) in the instantaneous
eigenstates, must satisfy the condition $\gamma({\bar r})\gg 1$.

The precession length $L$ coming from the effective Hamiltonian
(\ref{12}) is given by \cite{nunokawa}
 \[
 L=\frac{2\pi}{\sqrt{(2\mu_{ff'} B_\perp)^2+(V_{\nu_e}+\Omega_G-2\delta c_2)^2}}
 \]
At the resonance, it reads
 \[
 L_{res}=L({\hat r})=\frac{\pi}{\mu_{ff'} B_\perp}\simeq
 8\pi \,
 \frac{10^{-11}\mu_B}{\mu_{ff'}}\frac{B_c}{B_\perp}\,\mbox{m}\,,
 \]
The resonant spin flavor precession has the following width
 \[
 \delta r=2\Lambda \lambda\,,
 \]
where
 \[
 \Lambda =\left(\frac{\rho'({\bar r})}{\rho({\bar r})}\right)^{-1}
 =\frac{V_{\nu_e}({\bar r})}{V_{\nu_e}'({\bar r})}\,,
 \]
assuming
 \[
 y_f'(t, t)=0\,, \quad (Y_e'\ll n_e'/n_e)
 \]
and
 \[
 \lambda = 2\frac{l}{L_{res}}\,, \quad l=\frac{1}{2\delta}\,.
 \]
The spin flavor conversion is adiabatic provided
 \[
 \delta r \gg L_{res}\,,
 \]
which is equivalent to
 \[
\gamma = \frac{(\mu_{ff'} B)^2}{\delta  |\rho'/\rho|}\gg 1\,.
 \]
This condition can be immediately verified using the above profile
for $\rho(r)$, and $\mu_{ff'}\sim 10^{-11}\mu_B$ and the typical
values for the magnetic fields $B\sim (10^{12} - 10^{14})$G.

\subsection{Spin Flip Probability}

The conversion probability $P_{\nu_{fL}\to \nu_{f' R}}$ that the
neutrino left flips into neutrino right is
 \[
 P_{\nu_{fL}\to \nu_{f'R}}=\frac{1}{2}-\left(\frac{1}{2}-P_{LZ}\right)
 \cos 2{\tilde \theta}_i \cos 2 {\tilde \theta}_f\,,
 \]
where the Landau-Zener probability is given by
 \[
P_{LZ}=e^{-\gamma \pi/2}\,,
 \]
and the effective mixing angle ${\tilde \theta}$ is defined as
\cite{nunokawa,haxton}
 \[
\tan 2 {\tilde
\theta}(r)=\frac{2\mu_{ff'}B}{\Omega_G(r)+V_{\nu_e}-2\delta
c_2}\,.
 \]
${\tilde \theta}_i={\tilde \theta}(r_i)$ refers to initial mixing
angle at $r_i$ where neutrinos are produced (i.e. at the
neutrinosphere), and ${\tilde \theta}_f={\tilde \theta}(r_f)$ is
the mixing angle at $r_f$ where the neutrino helicity flip
probabilities are evaluated.

Finally, we note that the weak field approximation is fulfilled
since
 \[
 \frac{4GMR^2}{5{\bar r}^3}\, \omega R\lesssim 10^{-2}\,,
 \]
as $\omega\lesssim 10^4$Hz, and that rotational effects are
relevant during the time scale $t_0\lesssim$ 10 sec ($t_0$ is the
time scale for the emission of the energy $\sim 0.5\times
10^{53}$erg by each neutrinos degree of freedom with $p\sim
10$MeV) \cite{raffeltbook}.

\vspace{0.4in}

\section{Conclusions}

It has been suggested a mechanism for the generation of pulsar
kicks which accounts for the magnetic and gravitational fields of
a rotating nascent star. Owing to the relative orientation of
neutrino momenta with respect to the direction of the angular
velocity, the energy flux turns out to be generated
anisotropically. This effect is related to spin-gravity coupling,
which differs for neutrinos with opposite helicity (left- and
right-handed helicity). The estimations on $\Delta
m^2-\sin^2\theta$ parameters agree with the best fit of the
present data on neutrino oscillations.

Some conclusive remarks are in order:

\begin{itemize}
  \item  Results discussed in this contribution, as well as the papers
\cite{burrows,cuesta,loveridge}, in which pulsar kicks are
discussed in relation to gravitational waves, have been obtained
in semiclassical approximation, i.e. the gravitational field is
described by the classical field equations of General Relativity.
It will be of interest to investigate within the framework of
quantum theories of gravity.

  \item Results suggest a correlation between the motion of pulsars and
their angular velocities. Such a connection seems to be
corroborated by recent analysis and observations discussed in
\cite{laiOmega,wex}.

  \item The mechanism here proposed is strictly related to the
  gravito-magnetic effect, an effect predicted by General Relativity \cite{ciufolini},
  as well as by many metric theories \cite{will}. Its origin is due to the
  mass-energy currents (moving or rotating matter contribute to
  the gravitational fields, in analogy to the magnetic field of
  moving charges or magnetic dipole). Experiments involving the
  technology of laser ranged satellites \cite{ciufoliniPLA} are at
  the moment the favorite candidate to test gravitomagnetic
  effects.

  In connections with the mechanism proposed in this paper,
  a direct evidence of the gravitomagnetic effect seems to be provided by pulsar kicks.
  Future investigations on the velocity distribution of pulsars
  will certainly allow to clarify this still open issue.

\end{itemize}

\vspace{0.1in}

\section{Acknowledgments}

The author thanks the organizers, in particular Hans Thomas Elze,
of the Second International Workshop - DICE-04, {\it From
Decoherence and Emergent Classicality to Emergent Quantum
Mechanics}, September 1-4, 2004 - Piombino (Italy).

Many thanks also to A. Kusenko and J.F. Nieves for discussions.
Research supported by PRIN 2003.

\bibliography{apssamp}

\end{document}